\pdfoutput=1

\documentclass[reprint,showpacs,showkeys,preprintnumbers,amsmath,amssymb,aps,prl]{revtex4-1}

\usepackage{graphicx}
\usepackage{dcolumn}
\usepackage{bm}
\usepackage{amsmath}
\usepackage{amssymb}
\usepackage{latexsym}
\usepackage{epsfig}
\usepackage{amsbsy}
\usepackage{array}
\usepackage{amssymb}
\usepackage{setspace}
\usepackage{bm}

\def\sint{\ifmmode{- \!\!\!\!\!\! \int}
    \else{\hbox{$- \!\!\!\! \int \ $}}\fi}

\begin{document}

\preprint{Y. Q. Zhang et al, Proximity Effect of Spin Orbit Coupling}

\title{Proximity effect of spin orbit coupling in Pt/Co$_2$FeAl and Pt/Permalloy bilayers.}

\author{Yan-Qing Zhang}
\affiliation{School of Physics Science and Engineering, Tongji University, Shanghai 200092, China.}

\author{Niu-Yi Sun}
\affiliation{School of Physics Science and Engineering, Tongji University, Shanghai 200092, China.}

\author{Rong Shan}
\email{shanrong@tongji.edu.cn}
\affiliation{School of Physics Science and Engineering, Tongji University, Shanghai 200092, China.}

\author{Zhen-Gang Zhu}
\email{zgzhu@ucas.ac.cn}
\affiliation{School of Electronic, Electrical and Communication Engineering, University of Chinese Academy of Sciences, Beijing 100049, China.}

\date{\today}

\begin{abstract}

Proximity effect of spin orbit coupling is investigated through anomalous Hall effect in Pt/Co$_2$FeAl and Pt/Permalloy bilayers. A series of nontrivial magnetotransport behaviors, resulting from a strong impact of phonons on skew scattering, is observed in these films with ultrathin ferromagnetic layers. The parameters representing skew scattering, side jump and intrinsic contributions are dramatically enhanced when the ferromagnetic layer is very thin, and they have clear linear dependences on the reciprocal of ferromagnetic layer thickness, indicating a powerful influence of Pt/Ferromagnet interface. Further study on Cu/Co$_2$FeAl and Ta/Co$_2$FeAl bilayers reveals that a simple interface scattering without intense spin orbit coupling is not sufficient to trigger such a phenomenon. The proximity effect of spin orbit coupling is thus suggested to occur at the Pt/Ferromagnet interface, and as a result quite large anomalous Hall angle~(0.036) and Nernst angle~(0.23) are confirmed in the Pt/CFA films at room temperature.

\end{abstract}

\pacs{72.25.Mk, 72.25.Ba, 75.47.-m, 75.70.-i}

\keywords{magnetic film, proximity effect, spin orbit coupling, anomalous Hall effect, Hall angle}

\maketitle


The proximate thin Pt layer located at a Pt/Ferromagnet~(FM) interface can be polarized and has weak magnetic saturation magnetization~\cite{LCO2013}. This phenomenon, so called $'$magnetic proximity effect$'$~(MPE), attracts extensive attentions because it has possibly been involved in all spin-related transport properties such as the spin Hall effect in a system with Pt/FM (or using other metal instead of Pt) interface~\cite{HFQ2012,HLG2013,AMM2013,NAC2013,RRR2013,VSC2013}. In our opinion, since there is a clear action from the ferromagnet to the non-magnetic material, an opposite reaction must exist according to Newton's Third Law in a broad sense, especially in Pt/FM bilayers or multilayers. However, most studies are focused on features correlated to static magnetic properties, for example the high perpendicular magnetic anisotropy~\cite{WWS1994,SK1995,LFC1998,NKS1998,SSK1998,LRD2000,MTP2001,MRA2002,MRK2006,YSK2010}, while few reports mentioned the magnetotransport properties so far, e.g. the enhanced anisotropic magnetoresistance in these systems~\cite{Zhang1995,CLX2000,LCS2010,KAS2013}. The relevant research on this issue is seriously deficient. We thus employ the anomalous Hall effect (AHE) to investigate the influence of spin obit coupling at interfaces between Co$_2$FeAl (CFA), permalloy (Py, Fe$_{22}$Ni$_{78}$) and Pt layers systematically, because it is well known that AHE is strongly dependent on the intensity of spin orbit coupling in a magnetic object~\cite{KaLu1954,Smit1958,Berg1970,YKM2004,XCN2010,NSO2010}. Besides, Co$_2$FeAl is a magnetic Heusler alloy with distinguished features. The tunneling magnetoresistance (TMR) ratio of magnetic tunnel junction with Co$_2$FeAl electrodes can be over 330\% at room temperature~\cite{WSS2009}. DC pulse current-induced domain wall motion has been realized in Pt/CFA submicron wires with an ultrathin Co$_2$FeAl layer, which displays a similar performance to Pt/(Co/Ni)$_n$ multilayers~\cite{KS2013}. Co$_2$FeAl may be a very promising material for domain wall motion and spin transfer torque devices due to its larger TMR ratio for signal output. Our study would help to understand the spin transport behavior in this kind of structure.


A series of single CFA and Py layers with various thicknesses was deposited on 2.5-nm-thick Pt (or Cu, Ta) layer buffered MgO(100) substrates by using DC magnetron sputtering. For a comparison, MgO/CFA and MgO/CFA/Pt films were also prepared. MgO(100) substrates were baked at $700^\circ$C for an hour in the sputtering chamber before film growth. The CFA layer was sputtered from a stoichiometric Co-Fe-Al~(Co: 50\%, Fe: 25\%, Al: 25\%) target. The base pressure of the main chamber is around $2\times10^{-6}$~Pa. Ta layer in the sample was prepared at $700^\circ$C  and then CFA was deposited at $60^\circ$C , while all other samples were fabricated at $60^\circ$C. After the film growth process, all samples were annealed \textit{in situ} at $320^\circ$C for 35 minutes, which is an ideal thermal treatment temperature to induce the perpendicular magnetic anisotropy in Pt/CFA bilayers for device applications. The measurement of AHE was performed on the sample with Hall bar profile from 20~K to 300~K by a physical property measurement system (PPMS). The anomalous Nernst effect (ANE) measurement was carried out using a transport property measurement system designed by our group. The method to extract the anomalous Hall resistivity $\rho_{AH}$ from the original transverse resistivity $\rho_{xy}$ is the same as that in References~\cite{CSX2011} and~\cite{HMS2012}.

\begin{figure}
\centering
   \includegraphics[width=8.5cm]{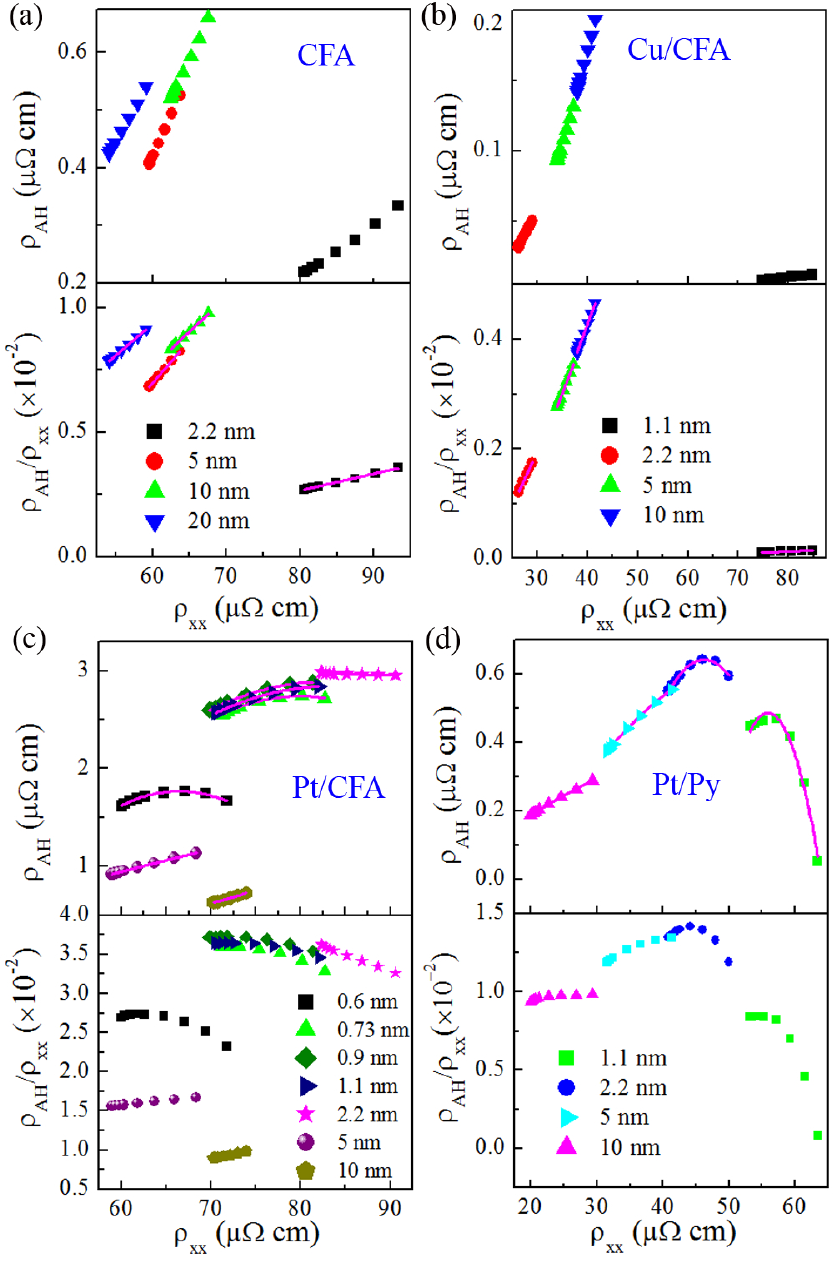}
   \caption{(Color online)  The dependences of $\rho_{AH}$ and $\rho_{AH}/\rho_{xx}$ on $\rho_{xx}$ are shown for CFA (a), Cu/CFA (b), Pt/CFA (c) and Pt/Py (d) films with various thicknesses. All magenta lines were fitted to $\rho_{AH}/\rho_{xx}$ data in (a) and (b), and to $\rho_{AH}$ data in (c) and (d).}
\label{fig:AHE}
\end{figure}

Two experienced formulas can be used to analyze AHE. One is a classical form with the expression of $\rho_{AH}=a\rho_{xx}+b\rho^2_{xx}$, where $\rho_{xx}$ is the longitudinal resistivity, $a$ represents the skew scattering parameter, and $b$ is dominated by the side jump and intrinsic contributions~\cite{KaLu1954,Smit1958,Berg1970}. Recently Tian Y et al. suggested the other expression $\rho_{AH}=a'\rho_{xx0}+a''\rho_{xxT}+b\rho^2_{xx}$, in which the skew scattering was divided into two parts: $a'$ resulting from the residual resistivity $\rho_{xx0}$ and $a''$ originating from the phonon-induced resistivity $\rho_{xxT}$~\cite{TYJ2009, SN2012}. The applicable circumstances of these two expressions were distinctly exhibited in our research. The dependences of $\rho_{AH}$ and $\rho_{AH}/\rho_{xx}$ on $\rho_{xx}$ are shown for CFA (a), Cu/CFA (b), Pt/CFA (c) and Pt/Py (d) films with various thicknesses in Fig.~\ref{fig:AHE}.  All magenta lines were fitted to $\rho_{AH}/\rho_{xx}$ data in (a) and (b), and to $\rho_{AH}$ data in (c) and (d). The films of CFA with thickness smaller than 1.1~nm are nonconducting because of the island growth on MgO substrates. The bilayer Cu/CFA with CFA layer thickness~($t_{CFA}$) around 0.6~nm met the similar problem too. The AHE in these two series of samples, nevertheless, can be handled by the classical expression once they are conducting as shown in Fig.~\ref{fig:AHE}(a) and (b). On the contrary, the new scaling must be employed when $t_{CFA}$ is less than 2.2~nm in Pt/CFA films, showing in Fig.~\ref{fig:AHE}(c). The situation is even more complicated in Pt/Py bilayers as shown in Fig.~\ref{fig:AHE}(d). Py is easier to be influenced by oxidization than CFA perhaps. The Pt/Py film with $t_{Py}=0.6$~nm is conducting, but it has a strange AHE behavior and cannot be fitted by both expressions (not shown here). The Pt/Py film with $t_{Py}=1.1$~nm can be fitted by the new scaling with a large error, while thick films can be fitted nicely like thick single Py layers using the same way~\cite{ZSS2013}, different from the Pt/CFA films.

A remarkable phenomenon is the increased Hall angle in the Pt buffered films, where we still use $\tan\theta=\rho_{AH}/\rho_{xx}$ as the Hall and Nernst angle although $\tan\theta\neq\theta$ in our experiments. Comparing with the thickest CFA film, a maximum enhancement of about 3.5 times in Hall angle is observed in Pt/CFA bilayers. The Hall angle of FePt, as we know, is very large, but it reduces by ~50\% from $20~K$ to $300~K$. The Hall angles of Pt/CFA bilayers hardly changes with varying temperatures. The Hall angles of the Pt/CFA bilayers with CFA thickness below 5~nm are actually larger than that of FePt film (20~nm) at room temperature~\cite{HMS2012}. Since the Hall angles of Py films are quite small ~\cite{ZSS2013}, they can be enhanced by the Pt buffer layer more forcefully.

\begin{figure}
\centering
   \includegraphics[width=7.5cm]{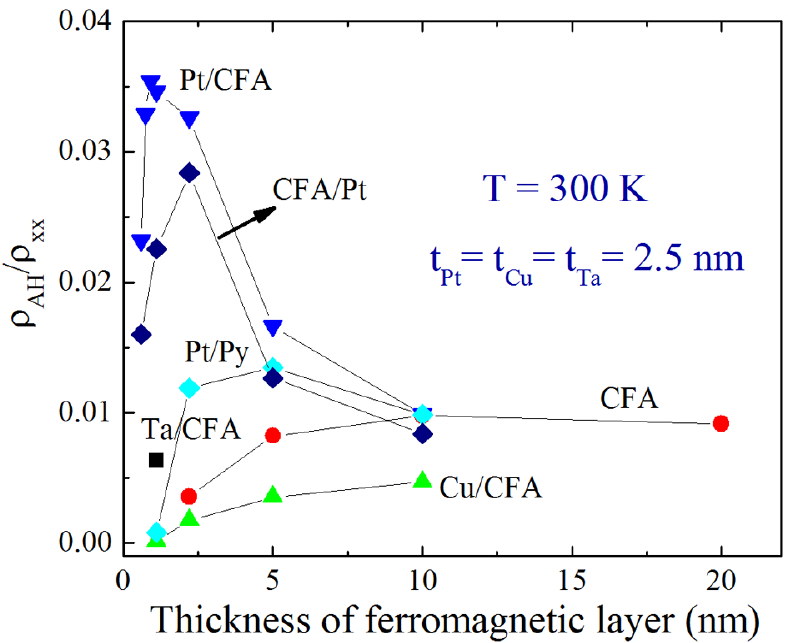}
   \caption{(Color online) Thickness dependences of the Hall angle~($\tan\theta=\rho_{AH}/\rho_{xx}$) for CFA, Pt/CFA, CFA/Pt, Cu/CFA, Ta/CFA and Pt/Py films at 300 K.}
\label{fig:AHA}
\end{figure}

Figure~\ref{fig:AHA} shows the thickness dependences of Hall angle for different films at room temperature. For the CFA and Cu/CFA films, the variation of $\rho_{AH}$ is slower than that of $\rho_{xx}$ from the interface scattering when CFA layer becomes thinner and thinner, so the Hall angles increase with increasing thickness and finally close to the intrinsic values of bulk material. The Cu/CFA bilayers have the lowest Hall angles among all samples because Cu layer has no significant contribution to $\rho_{AH}$ and the current was just bypassing through it. On the other hand, it is clearly illustrated that the interfacial scattering resulting from the thinning FM layer would not cause a large Hall angle in these bilayers. The enhancement of Hall angle has to be attributed to a strong spin orbit coupling at the interface in the Pt/CFA, CFA/Pt and Pt/Py films. 
A simple physical image is that moving conducting electron at the interface felt the interaction from the Pt layer are scattered in the CFA layer and this interfacial influence would not be suppressed too much by a finite thermal activation, which is the reason why the Hall angle does not change obviously with varying measurement temperature. 
The Hall angle enhancement cannot be owing to the MPE of Pt layer since its average moment is only $0.054~\mu_B$ at $300$~K~\cite{LCO2013}. Also, the results of the CFA/Pt bilayers exclude the influence of interface diffusion. The thinner CFA layer grown on MgO substrate, as we mentioned before, is not ideally continuous. Pt atoms should diffuse more easily into the CFA layer in this case. However, the Hall angle enhancement in CFA/Pt bilayers is weaker than that in Pt/CFA layers, indicating a favor of clean interface to the phenomenon. Combined with the data of Pt/Py bilayers, all results suggest that the enhancement is a magnifying effect rather than a Pt-diffused magnetic alloy effect. Additionally, the Ta layer deposited at $700^\circ$C  shows $\gamma$ phase, while it has $\beta$ phase at a lower growing temperature. The former exhibits a conventional temperature dependence of the resistivity as a normal metal, but the latter is opposite and gives rise to a strange AHE behavior~\cite{CMH1992}. Although it is difficult to compare Ta/CFA bilayer with other samples on AHE directly, it still exports a much larger Hall angle than the Cu/CFA bilayer with the same CFA thickness.

\begin{figure}
\centering
   \includegraphics[width=7.5cm]{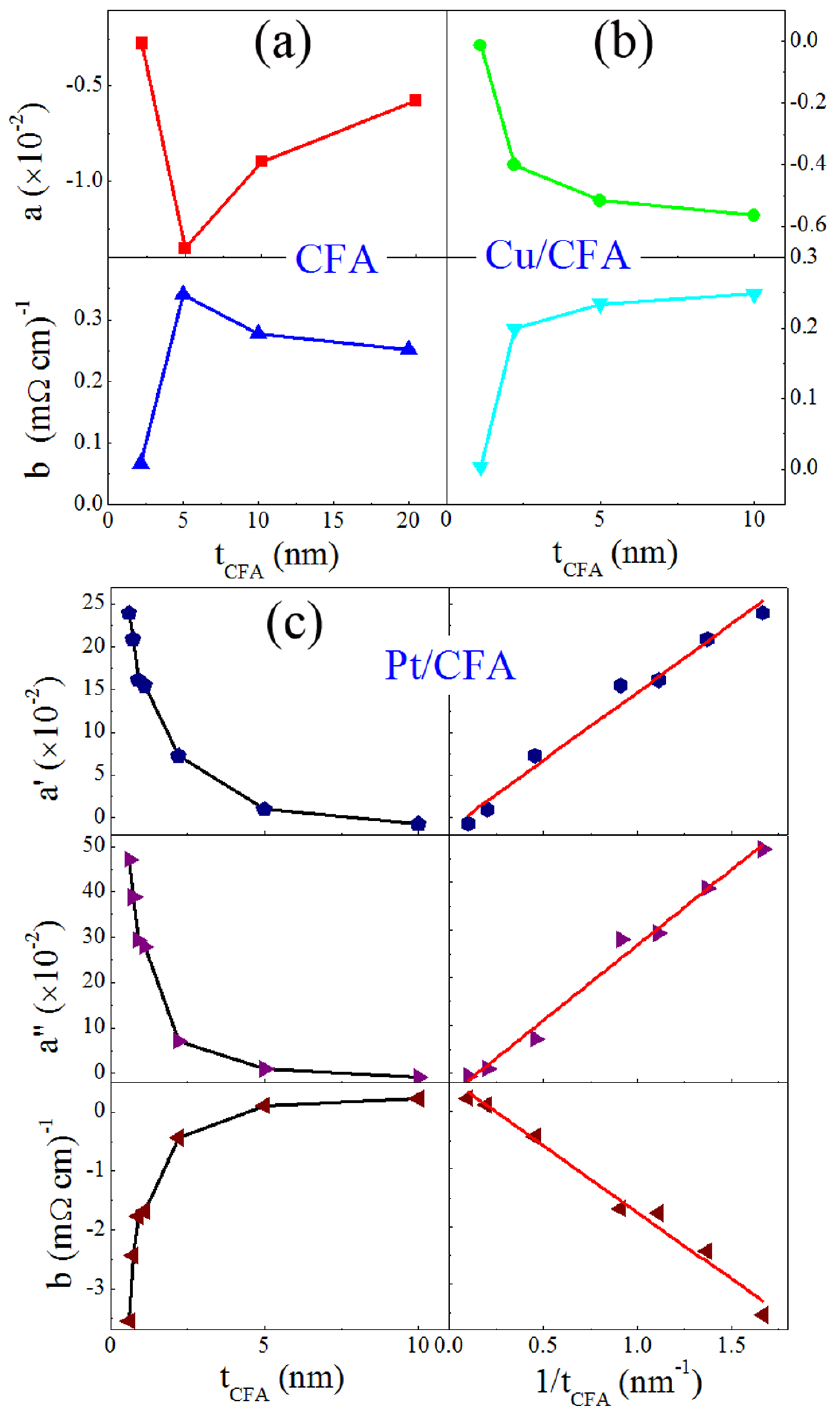}
   \caption{(Color online) Thickness dependences of $a$ ($a'$, $a''$) and $b$ for CFA (a), Cu/CFA (b) and Pt/CFA (c) films. The red solid lines in the right column of (c) are linearly fitted.}
\label{fig:abcurve}
\end{figure}

The fitted parameters of the AHE, which are dominated by the spin orbit coupling and the electronic structure of the magnetic layer in principle, disclose more details of the interfacial action. Usually due to the dimension effect in an ultrathin magnetic film, the conducting electrons are prevented from being scattered into up and down directions, and the skew scattering is enhanced consequently. In contrast, the reduction of integration of Berry curvature in k-space greatly restrains the intrinsic contribution to the AHE. Both mechanisms lead to typical trends of $a$ ($a'$, $a''$) and $b$ in the Fe and Py films with varying thicknesses~\cite{TYJ2009, ZSS2013}. Here the same law was observed in the CFA and Cu/CFA films, as shown in Fig.~\ref{fig:abcurve}(a) and (b). 
The factors $a$ ($a'$, $a''$) and $b$ of the Pt/CFA films are changing in the same way. However, their values are extremely larger than those of the CFA and Cu/CFA films with ultrathin CFA layers. Not only that, they are much larger than those of epitaxial Fe films on GaAs substrates in the same thickness range of the magnetic layer~\cite{TYJ2009}. It should be pointed out that the fittings are perfect for the AHE in these films and thus error bars are too narrow to be shown in the figures. Meanwhile, the signs of $a$ ($a'$, $a''$) and $b$ in the thinner Pt/CFA films are reversed comparing with those in CFA films, because the density of states of majority spins is cut down and more electrons with minority spins participate in the magneto-scattering at the interface. With the same reason, $a$ curve moves toward more positive value and $b$ curve moves toward more negative value in Cu/CFA films, but Cu interface cannot supply strong spin orbit coupling enough for the sign reversing. For the Ta/CFA (1.1~nm) bilayer, $a=0.06$ with a fitting error of $0.003$, while $b = -0.356$ with a fitting error of $0.0188$ in the unit of ({m$\Omega$~cm})$^{-1}$. The signs have already been reversed. In CFA/Pt bilayers the sign reversing occurred as well, meaning the chirality issue originated from the Dzyaloshinskii-Moriya interaction does not exist in this phenomenon~\cite{RTY2013}.

To sum up, we suggest a special region existing at the Pt/CFA, Pt/Py, Ta/CFA and Cu/CFA interface. In this region, the conducting electrons coming from nonmagnetic layer could take part in magneto-scattering, which leads to a peculiar AHE behavior. It is just like that conducting electrons in the magnetic layer side could feel the spin orbit coupling in the nonmagnetic layer, so we call it as $'$proximity effect of spin orbit coupling$'$. Different from the weak MPE, the intensity of this effect depends on the intrinsic property of the nonmagnetic layer. It could be very strong for the Pt/FM interface. This might be described by a surface Hamiltonian in analogy to a superconducting proximity effect to a topological insulator \cite{fu2008},
\begin{eqnarray}
\mathcal{H} &=& \frac{\mathbf{p}^{2}}{2m}+\frac{1}{4m^{2}c^{2}}\left(\frac{\mathbf{M}}{M_{s}}\times\nabla U\right)\cdot\mathbf{p} \notag\\
&+& \frac{1}{4m^{2}c^{2}}\left(\hat{\boldsymbol\sigma}\times\nabla V\right)\cdot\mathbf{p},
\label{Hamiltonian}
\end{eqnarray}
where the second term describes the spin-orbit effect in the magnetic layer with a magnetization \cite{KaLu1954}, the third term takes into account the effect of the Pt layer to a surface electron in the magnetic layer, $U(V)$ represents the periodic potential in ferromagnetic layer (Pt layer) and $\hat{\boldsymbol\sigma}$ are the Pauli matrixes. The surface spin orbit coupling stemming from the Pt layer may enhance the AHE signal in thinner magnetic layers since the weight of the surface contribution would be dominate in very thin films. However an observed tendency to zero AHE for zero thickness magnetic layer (Fig. \ref{fig:AHA}) is easily understood by noting that the AHE is vanishing for pure Pt layer. Therefore the Hall angle develops a broad peak for thinner magnetic layer as a consequence.

According to this physical image, a simple parallel connection model can be set up. $a$ ($a'$, $a''$) and $b$, all coding as $\eta$, should ascribe to the contributions of interface and bulk. Assume that the electron concentration ($n_e$) and the current density through the Pt/FM bilayers are uniform. Hence the measured parameters of AHE could be expressed as:
\begin{flalign}
&\mathcal{\eta}=\frac{n_edS}{n_et_{FM}S}\eta_I+\frac{n_e(t_{FM}-d)S}{n_et_{FM}S}\eta_b=\frac{d}{t_{FM}}(\eta_I-\eta_b)+\eta_b,&\nonumber\\\nonumber\\&\eta_b=\left\{\begin{aligned} 0,~~t_{FM}<d \\ \eta_{FM},~~t_{FM}>d \\ \end{aligned}  \right.&
\label{eq:LFunc}
\end{flalign}

\noindent where $S$ is the unit area in plane; $d$ is the thickness of the special region at the Pt/FM interface; $\eta_I$ and $\eta_{FM}$ is the intrinsic parameter of AHE of the special region and the FM layer, respectively. A linear relationship of $\eta$ and $1/t_{FM}$ can accordingly be deduced. The fitted red lines were shown in the right column of Fig.~\ref{fig:abcurve}(c), which is coincident with the simple model well and implies that d is samller than 6 \AA~at least. It is noted that the relationship must be modified when the Pt layer is too thick resulting in a large bypassing current.

\begin{figure}
\centering
   \includegraphics[width=8.5cm]{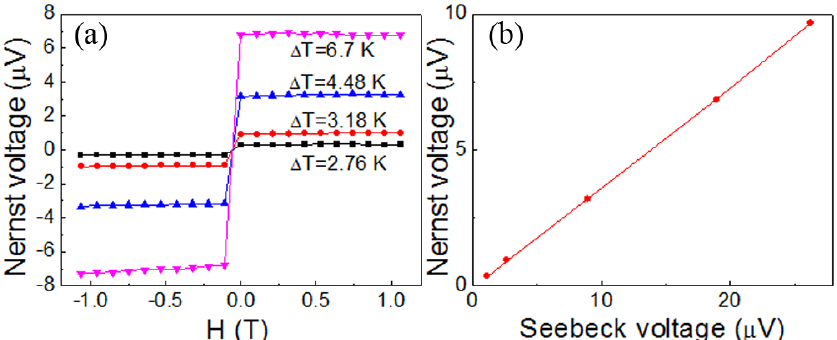}
   \caption{(Color online) Nernst hysteresis loops under different temperature gradients~(a) and a dependence of Nernst voltage vs. Seebeck voltage~(b), measured around room temperature.}
\label{fig:Nst}
\end{figure}

The proximity effect of spin orbit coupling induces novel characters in the Pt/CFA bilayers. For instance, the anomalous Nernst coefficient of a 80-nm-thick CFA film is about $9.5\times10^{-2}~\mu$V/KT~\cite{WAM2012}, while that of the Pt/CFA (1.1~nm) film is, nearly 9 times larger, around $0.83~\mu$V/KT. As a contrast, the Nernst coefficient of a 20-nm-thick FePt film measured in our system is $0.65~\mu$V/KT. Fig.~\ref{fig:Nst} shows the Nernst hysteresis loops under different temperature gradients~(a) and a dependence of Nernst voltage vs.~Seebeck voltage~(b), measured around room temperature. The sample has a robust perpendicular magnetic anisotropy even without any cap layer. Saturation magnetization of the sample is around 1000 emu/cc measured by Vibrating Sample Magnetometer (VSM). Similar to the enhancement of Hall angle, the anomalous Nernst angle of this sample, being 0.23, is also four times larger than that of the FePt film in Ref.~\cite{MOH2012}.


In conclusion, the proximity effect of spin orbit coupling at interface is demonstrated, which would have a strong influence on Pt buffered ultrathin magnetic films such as Pt/Co/AlO$_x$, Pt/(Co/Ni)$_n$, Pt/Py, Pt/CoFeB and Pt/CFA. All spin related transport properties, e.g. the spin transfer torque, are inevitably involved with this effect in these systems. What role the large Hall angle played is still an open question to the spin transfer torque and the spin injection in structures using Pt buffer or cap layer. New conceptions, based on this effect using an ultrathin ferromagnetic layer, might be ignited for spintronics devices.


\begin{acknowledgments}
This work was supported by the National Science Foundation of China Grant Nos. 51331004, 11374228, and the Innovation Program of Shanghai Municipal Education Commission No. 14ZZ038. Z. G. Zhu is supported by Hundred Talents Program of The Chinese Academy of Sciences. 
\end{acknowledgments}


%

\end{document}